\documentstyle[prl,aps,twocolumn,epsf]{revtex}

\catcode`\@=11

\def\maketitle2{\par 
\begingroup
\let\cite\@bylinecite
\def\thefootnote{\fnsymbol{footnote}}%
\twocolumn[\@maketitle2\vskip2pc]%
\thispagestyle{plain}\@thanks
\endgroup
\def\thefootnote{\arabic{footnote}}%
\setcounter{footnote}{0}%
\let\maketitle2\relax \let\@maketitle2\relax
\let\@thanks\relax \let\@authoraddress\relax \let\@title\relax
\let\@date\relax \let\thanks\relax \let\@abstract\relax 
\let\@pacs\relax}

\def\abstract#1{\gdef\@abstract{{\par 
\bgroup
\ifdim\prevdepth=-1000pt \prevdepth0pt\fi
\hsize\columnwidth
\dimen0=-\prevdepth \advance\dimen0 by17.5pt \nointerlineskip
\small\vrule width 0pt height\dimen0 \relax}{~~}#1\egroup}}

\def\pacs#1{\gdef\@pacs{{\par 
\bgroup
\hsize\columnwidth \parindent0pt
\ifdim\prevdepth=-1000pt \prevdepth0pt\fi
\dimen0=-\prevdepth \advance\dimen0 by20pt\nointerlineskip
\egroup} PACS numbers:~#1}}

\def\@maketitle2{
\@preprint
\@title
\ifdim\prevdepth=-1000pt \prevdepth0pt\fi
\@authoraddress
\@date
\begin{list}{}{\leftmargin=0.10753\textwidth \rightmargin=\leftmargin
\itemsep=1pc\partopsep=-1pc}
\item\@abstract
\item\@pacs
\end{list}
}

\catcode`\@=12

\def\bra#1{\langle#1|}		
\def\ket#1{|#1\rangle}		
\def\mod{\text{mod}}		
\makeatletter
\def\lsim{\compoundrel<\over\sim}
\def\compoundrel#1\over#2{\mathpalette\compoundreL{{#1}\over{#2}}}
\def\compoundreL#1#2{\compoundREL#1#2}
\def\compoundREL#1#2\over#3{\mathrel
  {\vcenter{\hbox{$\m@th\buildrel{#1#2}\over{#1#3}$}}}}
\makeatother

\begin{document}

\draft
\title{Quantum computation with phase drift errors.}

\author{C\'esar Miquel$^{1,2,3}$\thanks{miquel@df.uba.ar}, 
Juan Pablo Paz$^{1,2,3,4}$ \thanks{paz@df.uba.ar} and Wojciech Hubert Zurek
$^{1,3}$\thanks{whz@t6-serv.lanl.gov}}

\address{$^1$Institute for Theoretical Physics, University of California, 
Santa Barbara, CA 93106--4030, USA}

\address{$^2$Departamento de F\'{\i}sica ``J.J. Giambiagi'', 
FCEN, UBA, Pabell\'on 1, Ciudad Universitaria, 1428 Buenos Aires, Argentina}

\address{$^3$Theoretical Astrophysics, MSB288, Los Alamos National Laboratory,
Los Alamos, NM 87545, USA}

\address{$^4$Instituto de Astronom\'{\i}a y F\'{\i}sica del Espacio, 
CC67 Suc28, 1428~Buenos~Aires, Argentina}

\date{December 30, 1996}

\abstract{
We present results of numerical simulations of the evolution of an ion 
trap quantum computer made out of $18$ ions which are subject to a 
sequence of nearly $15000$ laser pulses in order to find 
the prime factors of $N=15$. We analyze the effect of random
and systematic phase drift errors arising from inaccuracies in the laser 
pulses which induce over (under) rotation of the quantum state. Simple 
analytic estimates of the tolerance for the quality 
of driving pulses are presented. We examine the use of watchdog 
stabilization to partially correct phase drift errors concluding that, in
the regime investigated, it is rather inefficient.
}


%
%

\maketitle2
\narrowtext

The key ingredient for quantum computation is the use of quantum parallelism
which takes advantage of the fact that the dimensionality of the Hilbert
space of the computer is exponentially dependent on its physical size. 
Feynman \cite{Feynman1} pointed out that 
classical computers are intrinsically inefficient when simulating the 
dynamics of a quantum system since for a chain of $L$ spins we need 
to store $O(2^L)$ complex amplitudes in memory. Since this quantity 
scales exponentially with the size, he reasoned, a quantum computer
would be much more efficient requiring, as nature does, only $O(L)$ 
sites. Shor's discovery \cite{Shor1} of a quantum algorithm for 
efficient factoring of integers is solid evidence that quantum computers 
could exponentially outperform their classical counterparts in problems 
which are not motivated by quantum mechanics {\it per se}. 
The last two years have witnessed an intense research effort aimed 
at examining the possibilities for taking quantum computation from the 
realm of ideas to the real world of the laboratory. However, 
practical implications of a ``quantum revolution'' for computation are 
still not clear. In this work we will analyze the performance of 
the ion--trap quantum information processor \cite{CZ1}, which is  
currently under study by various experimental groups around the world (see \cite{Steane}
for a review).
For concreteness we analyze the evolution of a quantum computer which 
runs a program to find the prime factors of a small number ($N=15$). 
Such program can be best represented by a quantum circuit 
decomposing into a complex sequence of elementary gates, i.e. 
unitary operations affecting qubits individually or in pairs \cite{Barenco}. 
In our simulations, which tested several factoring circuits
\cite{FACT}, we followed 
the quantum state of $n_i=18$ cold trapped ions subject to a 
predetermined sequence of $n_p\approx 15000$ 
(resonant and off--resonant) laser pulses.

These simulations face the same problem which 
prompted Feynman to propose the use of quantum computers in the first 
place: As their Hilbert space increases exponentially with their size, 
any dynamical study rapidly becomes a very hard computational task. 
Thus, to our knowledge, evolution corresponding to factoring in the 
proposed implementations of quantum computers has never been simulated 
beyond the level of a few qubits or small fragments of 
the complete algorithm \cite{CZ1}. Here, we present the first 
results of large numerical simulation of 
an ion trap quantum computer evolving under realistic (but still rather 
oversimplified) conditions.

We aim at examining the tolerance of quantum computation to 
errors which are likely to occur in any optically driven quantum--processor. 
In fact, for quantum gates to properly operate, 
most laser pulses are designed to invert population between internal levels 
($\pi$--pulses). However, the Rabi flopping frequency depends 
on a variety of physical effects which cause imperfections: 
Realistic $\pi$--pulses (or any other type of pulse) will 
always be $(\pi+\epsilon)$--pulses, being $\epsilon$ a random variable whose 
expectation value $\bar\epsilon$ and dispersion $\sigma$ characterize the 
quality of the experimental setup. In our study we analyzed the impact of 
these timing errors showing that a successful quantum computation (even on a rather 
small scale) imposes stringent limitations in the quality of laser pulses. Perhaps most 
importantly, we obtained and verified simple analytic expressions that 
could be used for estimating the fidelity of the final state of the computer. 
Our results make evident that error correction and fault tolerant computation
\cite{ErrCorr} are necessary for successfully implementing simple circuits. These techniques would make posible arbitrarily large computations
if the precision of the driving laser pulses are above a certain threshold. 
Unfortunately, simulating circuits which incorporate quantum error correcting codes 
is still beyond our capabilities: encoding a single qubit into $k$ (which should 
be at least 3) makes the time and memory requirements to grow by a factor 
$2^k$.

We have also set out to test the effectiveness of the watchdog (or quantum Zeno)
effect for error correction. The basic physics of this effect is
rather well known: Consider a two level system which is 
initially in state $|0\rangle$ and is subject to a sequence of
rotations by an angle $\theta$. After $k$ such rotations the 
probability for measuring the initial state is $P(0)=\cos^2(k\theta)$, 
which vanishes when $k\theta=\pi/2$. However, measuring the
state after each rotation tends to slow down evolution: the probability for finding 
the qubit always in state $|0\rangle$ is $P_w(0)=\cos^{2k}(\theta)$, which is  
close to one if $\theta$ is sufficiently small. 
At first sight, using this quantum Zeno effect to  
stabilize a quantum computation may seem implausible since to 
implement it one would have to know the ideal state of the computer at 
some times. However in an ideal factoring circuit, some of the qubits 
disentangle from the rest of the computer at predetermined steps of the
algorithm allowing for watchdog stabilization. 
With our simulations we tested this simple idea  
confirming the existence of watchdog stabilization but concluding that
the technique is rather inefficient in the ion trap quantum computer. 

In this implementation each qubit is stored in the internal levels of a 
single ion. Ions are linearly trapped and laser cooled to their translational
ground state (in the Lamb--Dicke limit). Two long-lived atomic ground 
states $\ket g$ and $\ket e$ of each ion play the role of the 
computational states (an additional auxiliary level $\ket{e'}$ is  
needed for implementing quantum gates). Each ion can be addressed by a 
laser and Rabi oscillations between the two computational states can be 
induced by tunning the laser frequency to the energy 
difference $\hbar \omega$ between ground and excited state. In this case, 
the quantum state of the qubit evolves as $\ket{\Psi(t)}=
U(t)\ket{\Psi(0)}$, where the matrix of $U(t)$ in the $(\ket{g},\ket{e})$
basis is:

\begin{equation}
U(t)= 
\left ( \begin{array}{cc}
  \cos \Omega t & -i e^{-i \Phi} \sin \Omega t \\
   -i e^{i \Phi} \sin \Omega t & \cos \Omega t \\
 \end{array} \right ) 
\label{vpulse}
\end{equation}

Thus, controlling the Rabi frequency $\Omega$, the laser phase $\Phi$ 
and the pulse duration $t$, arbitrary single qubit rotations 
can be performed. To implement two-bit gates one induces interactions 
between qubits using the center of mass mode as an intermediary:
Applying a laser pulse to ion $n$ with a frequency $\omega - \nu$, 
where $\hbar \nu$ is the energy of a single phonon of the CM mode, Rabi 
oscillations are induced between states $\ket g_{\text{n}} 
\ket{1}_{\text{\tiny CM}}$ 
and $\ket e_{\text{n}} \ket{0}_{\text{\tiny CM}}$. For these states 
the evolution operator is also (\ref{vpulse}), while states 
$\ket g_{\text{n}} \ket{0}_{\text{\tiny CM}}$ and $\ket e_{\text{n}} 
\ket{1}_{\text{\tiny CM}}$ remain unchanged by the interaction. 
These off-resonance pulses allow swapping information 
to and from the center of mass conditioned on the state of the 
$n$th qubit. As Cirac and Zoller showed \cite{CZ1}, by combining the two types 
of pulses applied on two different ions (and using an auxiliary level as a 
kind of ``work--space'') universal quantum gates 
can be implemented. Errors in $\Omega t$ and $\Phi$, 
such as the ones arising from fluctuations 
in the laser intensity which produce variations of $\Omega$, 
will result in over(under) rotations. Such {\it phase drift errors} 
are the ones of concern here. 
Other sources of errors, such us the decoherence of the CM mode or the 
spontaneous decay of the ions, 
will be ignored. In effect we assume that the computer 
evolves isolated from the environment being affected only by unitary errors.

The factoring circuit is based on Shor's 
algorithm \cite{Shor1}. Prime factors of $N$ are found 
by obtaining the order $r$ of a number $y$ which is coprime with $N$. 
This is the smallest integer such 
that $y^r=1 \ \mod \ N$ (with $r$ one computes the greatest common divisor 
between $N$ and $x^{r/2}-1$, which is a factor of $N$ whenever $r$ is even 
and $x^{r/2}\neq 1 \ \mod \ N$). To find $r$ one first chooses $y$ at random and
starts the computer in state
$|\Psi_0\rangle={1\over\sqrt{q}}\sum_{j=0}^{q-1}|j\rangle_1|0\rangle_2$. Here, 
$|j\rangle_{1,2}$ represent two registers of the computer 
whose states are defined by the 
binary representation of $j$ ($q$ must be between $N^2$ and $2N^2$). 
Then, by applying a unitary transformation mapping state 
$|j\rangle_1|0\rangle_2$ onto $|j\rangle_1|y^j \,\mod \,N\rangle_2$, the 
state of the computer is transformed into 
$|\Psi_1\rangle={1\over\sqrt{q}}\sum_{j=0}^{q-1}|j\rangle_1|y^j \,\mod \,
N\rangle_2$. 
Finally, one Fourier transforms the first register and measures it. The 
probability $P(c)$ for the outcomes of such measurement, whose  
analytic expression can be derived using standard quantum mechanics, is 
a strongly peaked distribution with the peaks separated by a multiple
of $1/r$. Thus, measuring the distance between peaks one efficiently gets  
information about $r$ (see Fig. 1.b). 

The most complicated part of the various factoring circuits 
\cite{FACT} is the modular exponentiation 
section (Fourier transform can be efficiently implemented 
as shown in \cite{Griffiths}). The 
complexity of the modular exponentiation network is 
such that the circuit involves $O(100)L^3$ elementary  
two bit gates \cite{FACT} 
(the two registers of the computer require $2L$ and $L$ qubits 
respectively). As unitary operators are invertible, the circuit must use 
reversible logic for which one needs a number of extra ``work--qubits'' 
in intermediate steps of the calculation. 
For the simplest circuits, such number is $2L+1$. 
However, other networks 
reduce the size of the workspace enlarging the number of operations. For 
example, another circuit we investigated has $L+1$ work--qubits but 
requires $O(10)L^5$ elementary operations (notably, for small numbers 
this circuit outperforms all others both in space and time). 
We will not discuss any circuit details here. For the purpose 
of analising the physical constraints implied by efficient factoring 
on the accuracy of laser pulses it is sufficient
to say that the simulations were performed on circuits involving
the following characteristics: 18 two level ions were 
subjected to 15000 laser pulses ($\sim 10^4$ off--resonant and 
$\sim 5 \times 10^3$ resonant). Eight of these ions were used in the first 
register, four in the second and six were used as work--qubits.
If all degrees of freedom are taken into account the Hilbert space of the computer
is $2\times3^{18}$ dimensional (each ion contributes with three levels and the CM is
effectively two dimensional). Unfortunately, this is too much for a classical computer.
However, if we consider all pulses involving the auxiliary level $\ket{e'}$ of 
each ion as perfect a substantial saving 
is achieved. In this case, the effective
dimension of the Hilbert space is $2^{19}$ which allows for numerical simulations. 
Taking this into account, the erroneous pulses amount to 75\% - 80\% of the total. 
The errors in $\Omega t$ and $\Phi$ were taken as random variables normally distributed 
with dispersion $\sigma$ and mean $\bar \epsilon$.

\begin{figure}
\epsfxsize=8.6cm
\epsfbox{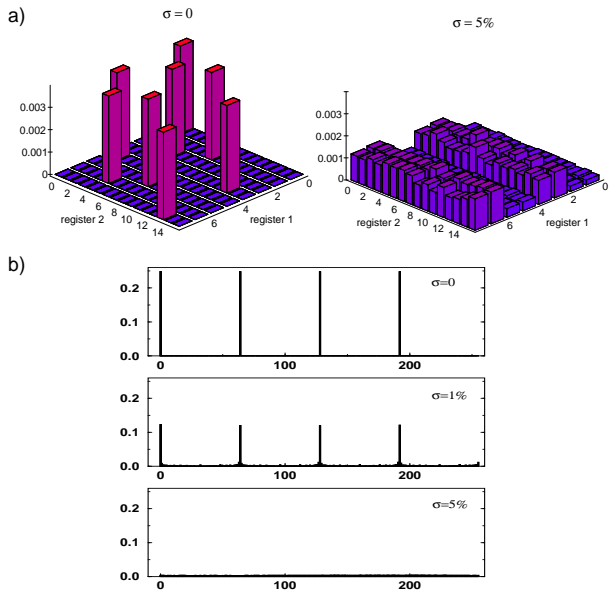}
\caption{a) Probability for a measurement of the two registers of a 
factoring computer. 
After an error--free modular exponentiation circuit, the computer should be 
in a superposition of the form  $\ket j \ket{y^j \,\mod \,N}$. As $y=7$, the only 
states present should be: $\ket 0 \ket 1$, $\ket 1 \ket{7}$, $\ket 2 \ket{4}$, 
etc (left figure). Timing errors distributed with dispersion $\sigma=5\%$ 
produce a probability showing no resemblance to the ideal one. 
b) Probability of measuring a value in the first register 
after Fourier transform for $\sigma=0,\ 1$ and $5\%$. 
For the last case all periodicity is lost 
and a uniform distribution arises. If $\sigma \lsim 1\%$ 
the distribution shows minor changes in amplitude.}
\label{figure1}
\end{figure}

An illustrative example of the data we computed can be seen in Fig. 1 where we 
represent the joint probability for measuring the two registers of the 
computer in values $(a,x)$ before the Fourier transform is calculated
(Fig. 1.b shows the distribution after Fourier transform is applied).
For the case of no systematic error ($\bar\epsilon =0$) it is clear that a 
dispersion of $5\%$ in the accuracy of the pulses 
completely wipes out the signal one wants to observe (note that 
the signal disappears before applying the FT circuit. As
discussed in \cite{CZ1} the FT circuit is quite resistant to this
level of errors but our result shows that, unfortunately, this is not the case
for modular exponentiation where many more operations are needed). A reasonable parameter 
quantifying the accuracy of the quantum computer is the fidelity, which is 
defined as the overlap between the actual state (obtained after evolution 
subject to noise) and the ideal one: $F=|\bra{\Psi_{\text{actual}}}
\Psi_{\text{ideal}}\rangle|^2$. 
Our observations show that a fidelity below $1/5$ implies the loss of 
the signal one wants to observe (in general $F$ is closely related to
the probability of observing the system in the correct state, i.e.
measuring the first register on a peak in Fig. 1b). In Fig. 2 we show 
the dependence
of fidelity upon the dispersion of the errors. The numerical results
follow remarkably well a simple formula that can be derived by 
assuming one has $l$ {\sl independent} qubits each one of which is 
subject to $n/l$ erroneous pulses. In this way (treating the center of 
mass motion separately, as it is subject to a larger number of 
pulses) we conclude that the mean 
fidelity (averaging over the ensemble of errors) is  
\begin{equation}
\bar F=\left[{1\over 2}\left(1+e^{-2 n_{\text{t}} \sigma^2/l} \right ) \right ]^l \ \left [ {1\over 2} \left ( 1 + e^{-2 \sigma^2 n_{\text{cm}}} \right ) \right ], \label{fidelity}
\end{equation}
where $n_{\text{t}}$ is the total number of pulses and $n_{\text{cm}}$ is 
the number of off-resonance ones (which involve the center of mass). 
This rough estimate was also used to estimate the dependence of the fidelity on the 
number of operations (for fixed dispersion $\sigma$) giving also good 
quantitative agreement with the simulations. 
We numerically computed a fidelity for every realization of the noise finding the
average over many (i.e. a few tens) noise realizations. Error bars in 
Figure 2 correspond to the dispersion around the mean of the numerically computed 
result. Fortunately, fluctuations are relatively small and therefore each run gives
a reasonable idea of the average result. The reason for this is that each run 
corresponds to a random choice of many (nearly $15000$) independent random variables. 
Therefore, fluctuations between different runs are effectively suppressed.

It is also interesting to estimate the number of dimensions explored by the 
state of the computer, which while moving on a large Hilbert space is subject
to random perturbations. For this purpose we computed the entropy of the 
density matrix $\rho_{\text{av}}$, obtained by averaging the state vector on the ensemble 
of noise realizations. Linear entropy $S_{\text{lin}}=-\log_2(\,\text{Tr}\,\rho_{
\text{av}}^2 \,)$ (which 
provides a simple lower bound to von--Neumann entropy) turns out to be 
well approximated by:
\begin{equation}
S_{\text{lin}}=l+1-\log_2\left[(1+e^{-4n_{\text{t}}\sigma^2/l})^l(1+e^{-4n_{\text{cm}}\sigma^2})\right].
\end{equation}
This equation was shown to agree with numerical results and predicts that 
for dispersions above a few percents the computer explores the entire available 
Hilbert space (enough statistics to numerically test the above formula was only 
gathered for dispersions below $2\%$).

\begin{figure}
\epsfxsize=8.0cm
\epsfbox{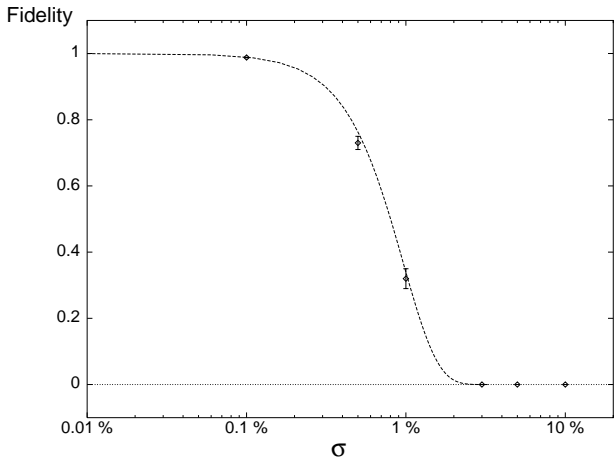}
\caption{Fidelity as a function of the noise dispersion $\sigma$ 
(for $\bar\epsilon=0$) at the end of the modular exponentiation circuit 
for factoring $N=15$ (with $q=256$ and $y=7$). The dashed line is the na\"{\i}ve
qualitative estimate given by equation (2) with $n_{\text{t}}=1.5 \times 10^4$, $n_{\text{cm}}=10^4$.}
\label{figure2}
\end{figure}

To analyze the effectiveness of watchdog stabilization we 
simulated a smaller version of our factoring $N=15$ circuit with only the first 
three controlled multipliers. In these simulations we assumed a systematic error using
$\bar\epsilon = 1.1 \sigma$. Whenever a work--qubit was expected to be 
in a state $\ket 0$ a measurement was performed on it. As the center of mass is
supposed to return to the ground state after each gate, a measurement of its
state was also carried out (in practice this could be done using a ``red
light ion'' as suggested in \cite{CZScience}). The probability for finding all these 
qubits in the correct state was recorded and the computation was continued on the 
correct branches only. 
The final fidelity, defined as the 
probability for the sequence of correct results, was computed and compared with the 
previously described one (when no watchdog was performed). 
The simulations show that watchdog stabilization produces a minor improvement 
in the fidelity: For example, when $\sigma = 0.1\%$ the fidelity with 
watchdog was 0.67 as opposed to 0.64 without it. To estimate the improvement 
one should expect we computed the fidelity of $l$ independent qubits 
which are subject to watchdog stabilization after each of $n/l$ rotations 
by an angle $\bar\epsilon$. In this case, the result is rather surprising: this 
ideal watchdog would produce a fidelity close to $0.99$, a number which is much
larger than the one computed numerically. 
This inefficiency of the stabilization method can be explained as follows:
Measuring the state of a few qubits (the ones that disentangle 
from the rest of the computer at some intermediate times) 
one is doing a ``partial watchdog". Thus, after the measurement we know with 
certainty that the measured qubits are in the ``correct" state but 
the rest of the computer may be in an erroneous one. To test this simple 
explanation we run a numerical simulation where,  
after each measurement, the state of the computer was projected onto the 
ideal one. In this case, the agreement with the na\"{\i}ve  (independent qubit) 
estimate was good being the fidelity close to $0.99$. 

One of the interesting results of our study is that, although the factoring 
circuit continuously correlates the qubits, the dependence of fidelity on 
the noise parameters, can be estimated using a simple model where 
non--systematic errors affect each qubit independently (with this model, 
equation (\ref{fidelity}) can be easily obtained). However, 
for purely systematic errors ($\sigma=0$) we were not 
able to obtain a simple analytic estimate for the fidelity fitting the 
results of our simulations. For example, assuming that every qubit evolves 
independently under the influence of $n/l$ pulses which produce a rotation 
in an angle $\bar \epsilon$ one gets a formula for the mean fidelity which differs 
from eq. (\ref{fidelity}) by a term $\cos 2 \bar\epsilon n/l$ multiplying 
each exponential. 
This na\"{\i}ve estimate predicts
a lower fidelity than the one we numerically computed. The reason for this seems
to be the existence of cancellations of errors associated, in a nontrivial way, 
with the reversible nature of the circuit (for example, in a controlled not gate 
systematic errors exactly cancel when the control qubit is in the ground state but
propagate otherwise). This suggests that pulse sequences implementing 
logic gates should be designed to properly compensate for systematic 
over(under) rotations. If this is achieved, the remaining fidelity is well 
approximated by equation (\ref{fidelity}). 

We acknowledge the hospitality of the ITP at Santa Barbara
where this work was completed. This research was supported in part
by the NSF grant No. PHY94--07194. JPP was also supported by grants 
from UBACyT, Fundaci\'on Antorchas and Conicet (Argentina).

%
%

%
%
\end{document}